\documentclass[onecolumn]{aastex6}

\collaborationName{Friends of AASTeX}

\begin{document}


\title{On the claimed X-shaped structure in the Milky Way bulge}


\author{Daniel Han and Young-Wook Lee}
\affil{Center for Galaxy Evolution Research and Department of Astronomy, Yonsei University, Seoul 03722, Korea \\
\url{daniel.han@yonsei.ac.kr}, \url{ywlee2@yonsei.ac.kr}}





\begin{abstract}
A number of recent studies have claimed that the double red clump (RC) observed in the Milky Way bulge is a consequence of a giant X-shaped structure. In particular, \citet{2016AJ...152...14} reported a direct detection of a faint X-shaped structure from the residual map of the Wide-Field Infrared Survey Explorer (\textit{WISE}) bulge image. Here we show, however, that their result is affected substantially by whether the conventional dust extinction correction is applied or not and partly by a bulge model subtracted from the original image. We find that the residuals obtained by subtracting either ellipsoidal or boxy bulge models from the dereddened images show no obvious X-shaped structure. We further show that, even if it is real, the stellar density in the claimed X-shaped structure is way too low to be observed as a strong double RC at $l = 0\degr$.
\end{abstract}




\section{Introduction} \label{sec:intro}

Starting with the discovery of the two RCs in the bulge color-magnitude diagram, it is widely believed that most of the stellar component of the Milky Way bulge is in a giant X-shaped structure \citep{2010ApJ...724...1491,2010ApJL...721...L28,2013MNRAS...435...1874}. This discovery, together with the dynamical simulations where the X-shaped structure can naturally arise via bar buckling \citep{2012ApJL...757...L7}, has led to the suggestion that the ``pseudo bulge'' is not restricted to the low-latitude region of the bulge, but expanded even to the high-latitude field. However, a completely different interpretation has been suggested by \citet{2015MNRAS...453...3906} and \citet{2017ApJ...840...98}, according to which the double RC phenomenon is caused by two stellar populations with different helium abundances. Since the helium-enriched stars are brighter than helium-normal stars \citep{2005ApJ...621...L57,2010ApJL...715...L63}, this can reproduce the two RCs without a difference in distance.

\begin{figure}[ht!]
\figurenum{1}
\plotone{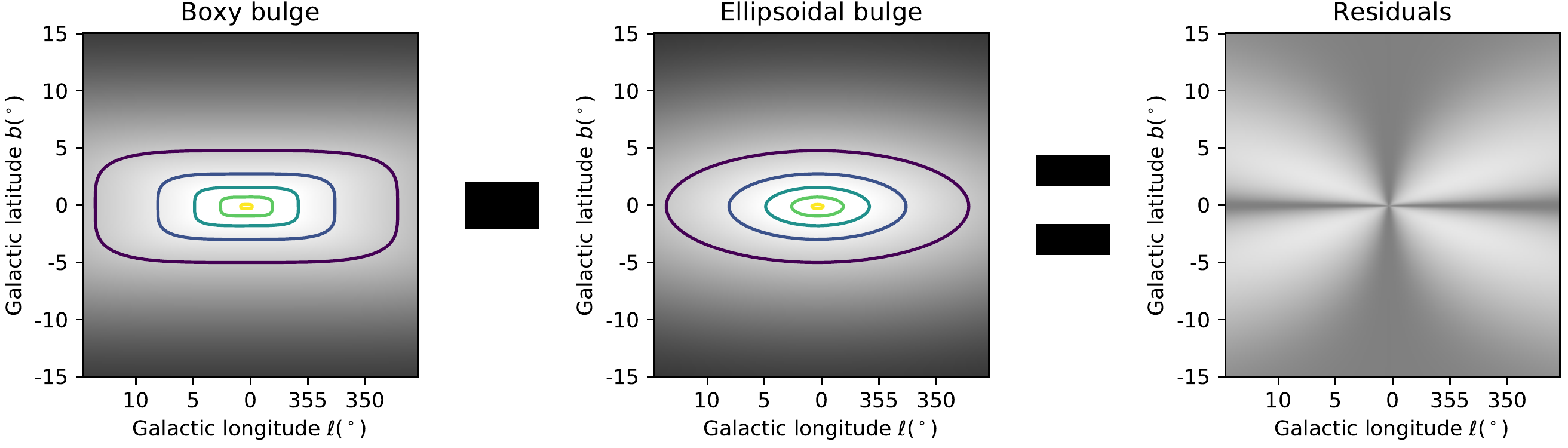}
\caption{Formation of an artificial X-shaped structure. When an ellipsoidal bulge model (middle panel) is subtracted from a boxy bulge model (left panel), an artificial X-shaped structure remains in the residual map (right panel) \label{fig:toymodel}}
\end{figure}

Recently, \citet[hereafter NL16]{2016AJ...152...14} claimed that an X-shaped structure is revealed when a simple exponential model is subtracted from the \textit{WISE} image of the Milky Way bulge (see their Fig. 3). However, there are a couple of potential problems in their analysis. One is a lack of appropriate correction of dust extinction which is non-negligible even in the infrared image of the bulge. The other is the use of a simple ellipsoidal model for the bulge, while it is well-known that the Milky Way has a boxy bulge \citep{1995ApJ...445...716,2005AA...439...107,2011AA...534...A3}. As illustrated in a toy model in Figure \ref{fig:toymodel}, an artificial X-shaped structure remains in the residual map when an ellipsoidal bulge model is subtracted from a boxy bulge. This effect is well-known in the extragalactic community, for example, the similar artificial X-shaped structures appear when ellipse-fit galaxy models are subtracted from the images of early-type galaxies \citep[see e.g.,][]{2011ApJ...730...23,2016ApJ...822...95}. The purpose of this paper is to investigate the effects of the adopted bulge model subtracted from the original image and the dust extinction correction on the residual maps.

\section{Extinction corrections and Model fitting} \label{sec:method}

\begin{figure}[h!]
\figurenum{2}
\epsscale{0.6}
\plotone{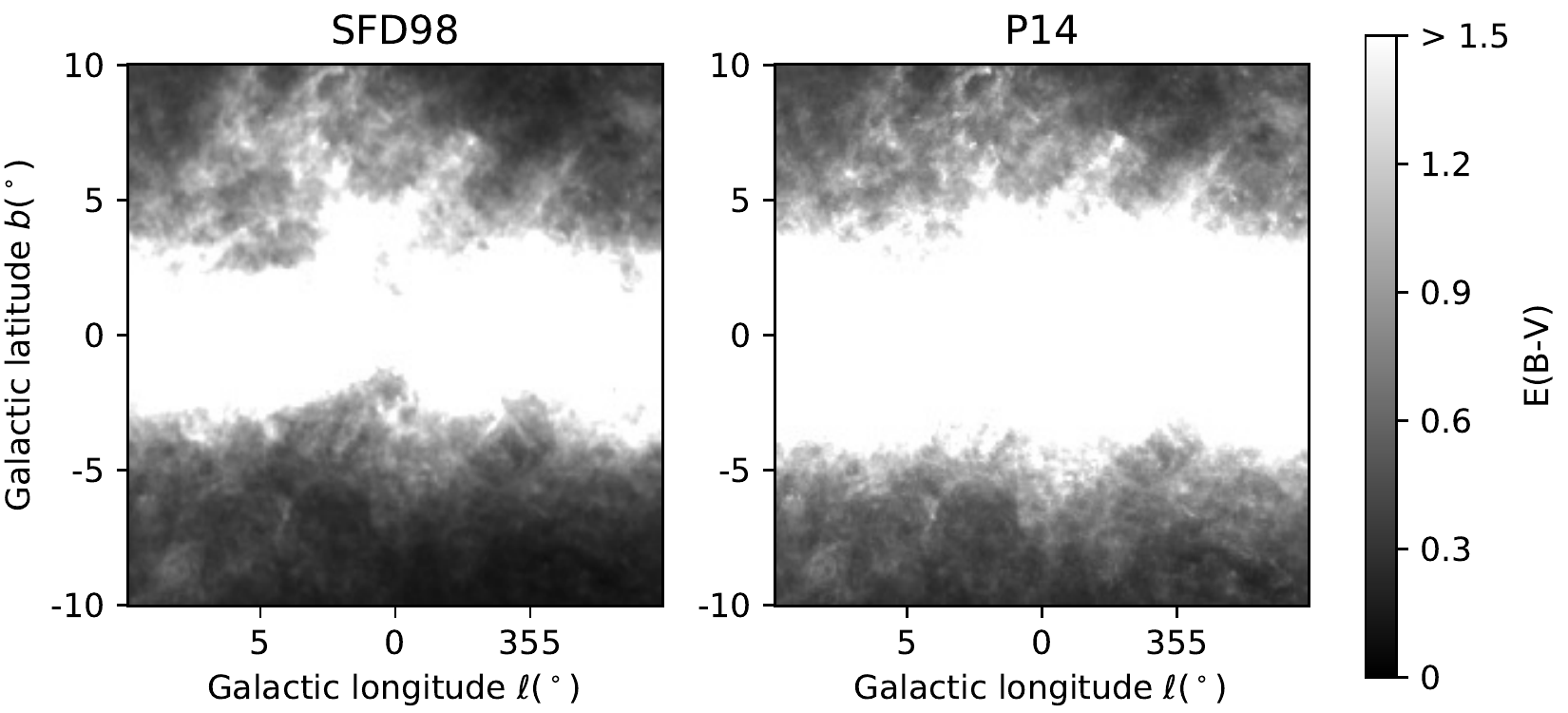}
\caption{E(B-V) maps of \citet[left]{1998ApJ...500...525} and \citet[right]{2014AA...571...A11}. Note that E(B-V) values at very low-latitude regions ($|b| < 2\degr$) well exceed the maximum value of the scalebar. \label{fig:dustmaps}}
\end{figure}

Following NL16, we use the ``unWISE'' images of the bulge fields across $(|l|, |b|) < (10\degr, 10\degr)$ in the \textit{WISE} \textit{W1} and \textit{W2} bands. A detailed procedure for the construction of ``unWISE'' images is described in \citet{2014AJ...147...108}. In order to investigate the effects of dust extinction on the residual images, we choose two dust maps (Fig. \ref{fig:dustmaps}), one of which is the most widely adopted by the community \citep[hereafter SFD98]{1998ApJ...500...525}, and the other provides the highest resolution \citep[hereafter P14]{2014AA...571...A11}. The resolutions of the two maps are $6\arcmin$ and $5\arcmin$, respectively, which are equivalent to $\sim 4$ pixel in the bulge images we employed. Despite the progress in the construction of the dust maps during the past decades, the uncertainty in the reddening correction still increases rapidly toward the Galactic plane. Therefore, the low-latitude zones $(|b| < 2\degr)$ are masked in our analysis as has been done in other recent studies \citep[e.g.,][]{2013AA...552...A110,2016PASA...33...e025}. The pixels in the top $5\%$ and bottom $5\%$ in the \textit{W1}-\textit{W2} color distribution are also masked as in NL16.

To see the effects of the bulge models on the residual images, first, we use the same ellipsoidal bulge model employed in NL16, which has a simple exponential profile with an axis ratio of $0.38$. We then employ the triaxial boxy bulge model of \citet[hereafter L05]{2005AA...439...107}, which is based on the 2MASS star counts. The axial ratios of this boxy bulge model are $1:0.49:0.37$ and the major axis is oriented with respect to the Sun-Galactic center direction of $29\degr$. The normalization coefficient in the bulge model fitting is determined by the chi-square minimization:
\begin{displaymath}
\sum_{(x, y)}\chi^2(a) = [\frac{D(x, y) I(x, y) - a B(x, y)}{\sigma(x, y)}]^2,
\end{displaymath}
with $D(x, y) = 100^{\frac{2}{5}(R(W_1, W_2) E(B-V))},$ where:

\noindent $D(x, y)$ is the dust extinction correction term for each pixel;

\noindent $I(x, y)$ is the pixel value at each pixel in the bulge image;

\noindent $a$ is the normalization coefficient between the bulge image and model;

\noindent $B(x, y)$ is the pixel value at each pixel in the bulge model;

\noindent $\sigma(x, y)$ is the Poisson noise at each pixel in the bulge image;

\noindent $R(W_1, W_2)$ is the extinction in the \textit{W1} or \textit{W2} band relative to E(B-V), from \citet{2013MNRAS...430...2188}

\newpage
\section{Residual images of the bulge} \label{sec:results}
\begin{figure}[h!]
\figurenum{3}
\plotone{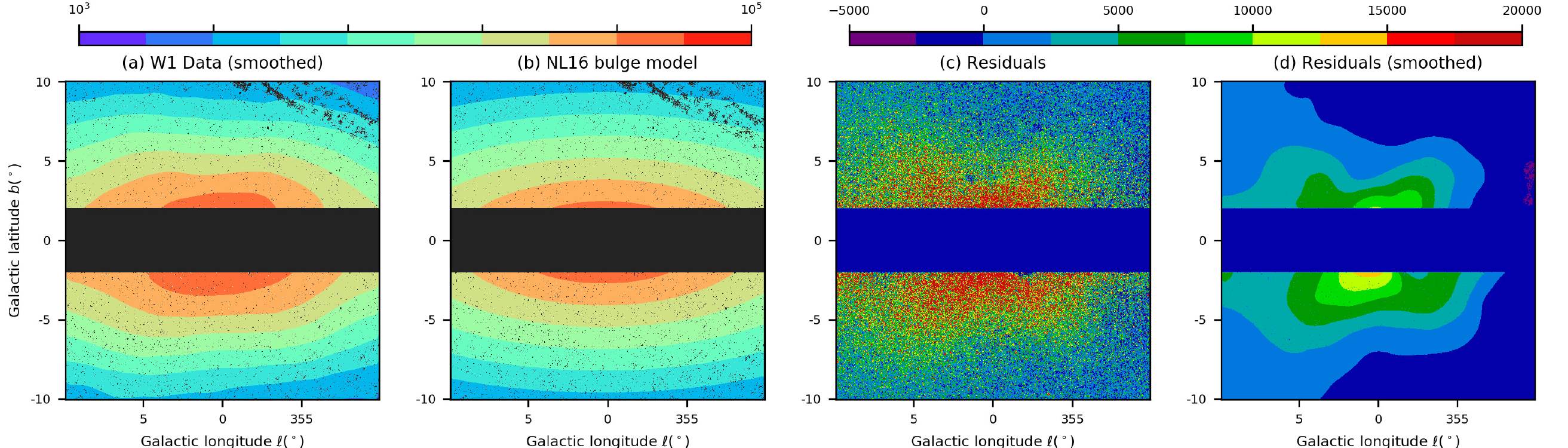}
\caption{\textit{WISE W1} data (panel (a)) fit by bulge model of NL16 (panel (b)) before the dust extinction correction. Panels (c) and (d) are residual maps. As has been done in NL16, 25 pixel median smooth filtering is applied in panels (a) and (d) to see the overall structure rather than particular details. \label{fig:original-NL16}}
\end{figure}

\begin{figure}[h!]
\figurenum{4}
\plotone{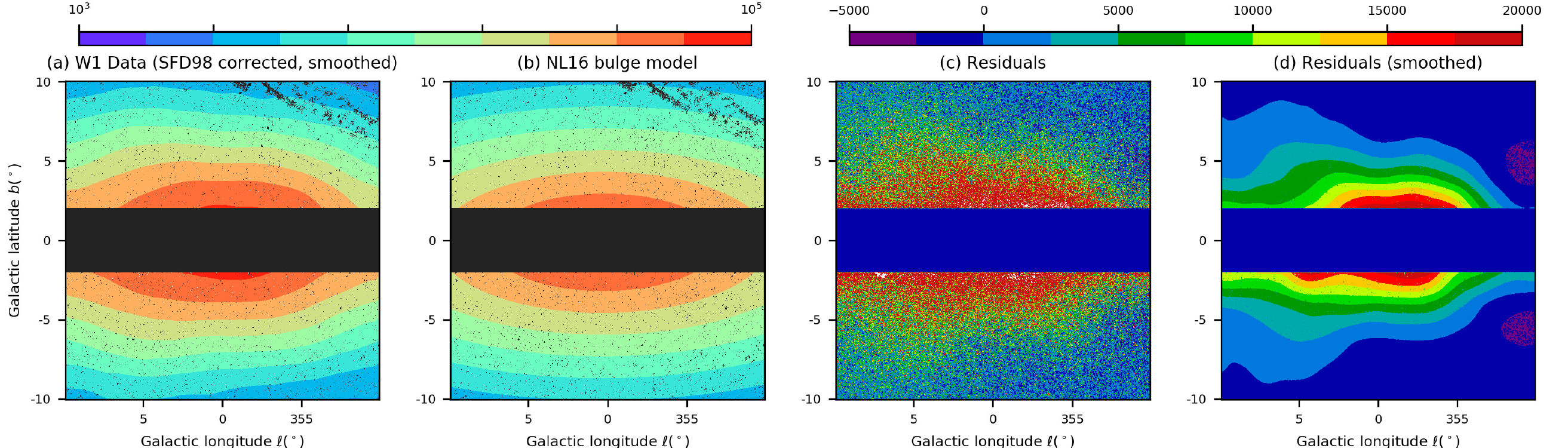}
\caption{Same as Fig. \ref{fig:original-NL16}, but after the dust extinction correction with the E(B-V) map of SFD98. \label{fig:sfd98-NL16}}
\end{figure}

\begin{figure}[h!]
\figurenum{5}
\plotone{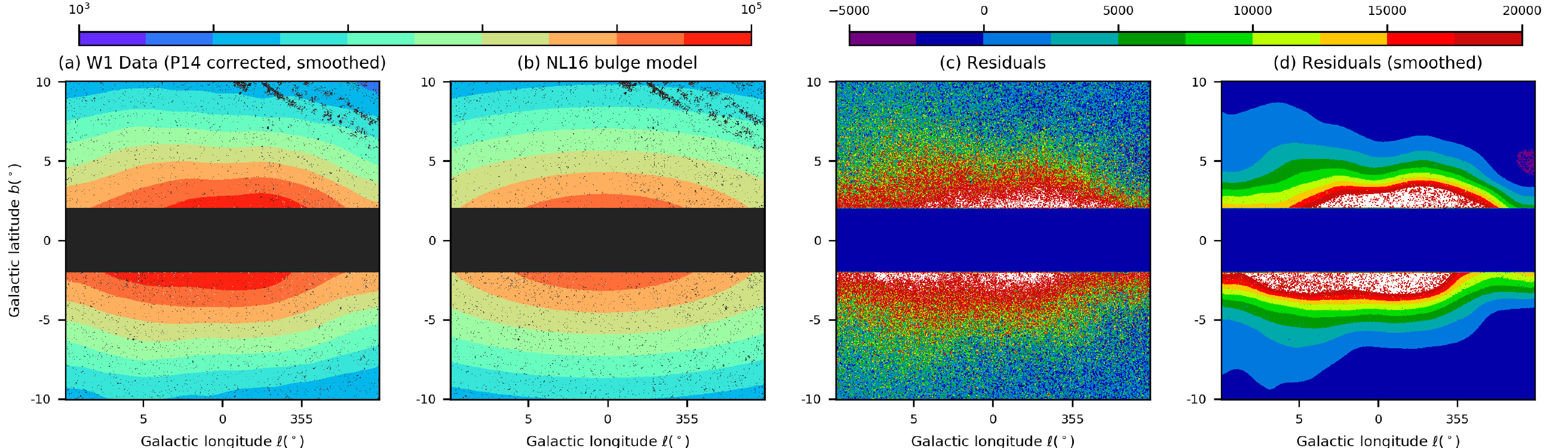}
\caption{Same as Fig. \ref{fig:sfd98-NL16}, but with the E(B-V) map of P14. \label{fig:planck-NL16}}
\end{figure}

Before we proceed to the effects of dust extinction correction, Figure \ref{fig:original-NL16} shows the residual maps with the uncorrected \textit{WISE W1} data which is essentially identical with those in NL16 (see their Fig. 3). Because we mask the low-latitude regions, here we show the images in the logarithmic and linear scales, while NL16 used arcsinh scale to highlight the faint X-shaped structure in the image without masking the extremely bright central region. The asymmetry of the images in Figure \ref{fig:original-NL16} is due to the effect of perspective of the elongated bar projected on the sky. The effects of dust extinction corrections are then shown in Figures \ref{fig:sfd98-NL16} and \ref{fig:planck-NL16} for the two different E(B-V) maps. As is clear from these figures, after the dust extinction corrections, a boxy structure, rather than the X-shaped structure, is left in the residual maps. This suggests that the result of NL16 was noticeably influenced by a lack of the dust extinction correction.

\begin{figure}[h!]
\figurenum{6}
\plotone{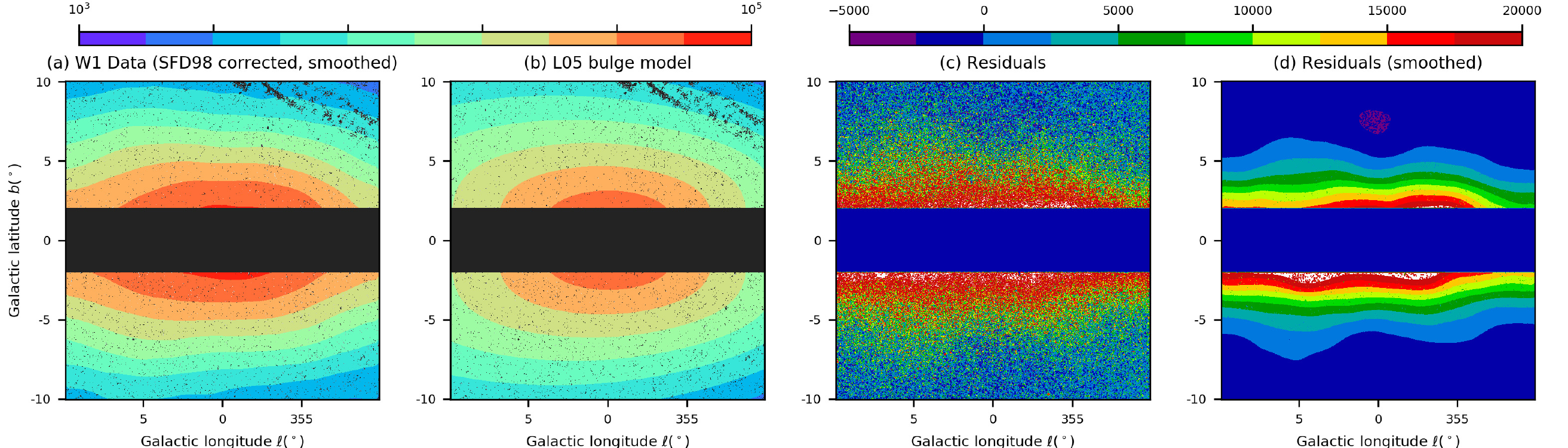}
\caption{Same as Fig. \ref{fig:sfd98-NL16}, but after the subtraction of the L05 boxy bulge model. \label{fig:sfd98-L05}}
\end{figure}

\begin{figure}[h!]
\figurenum{7}
\plotone{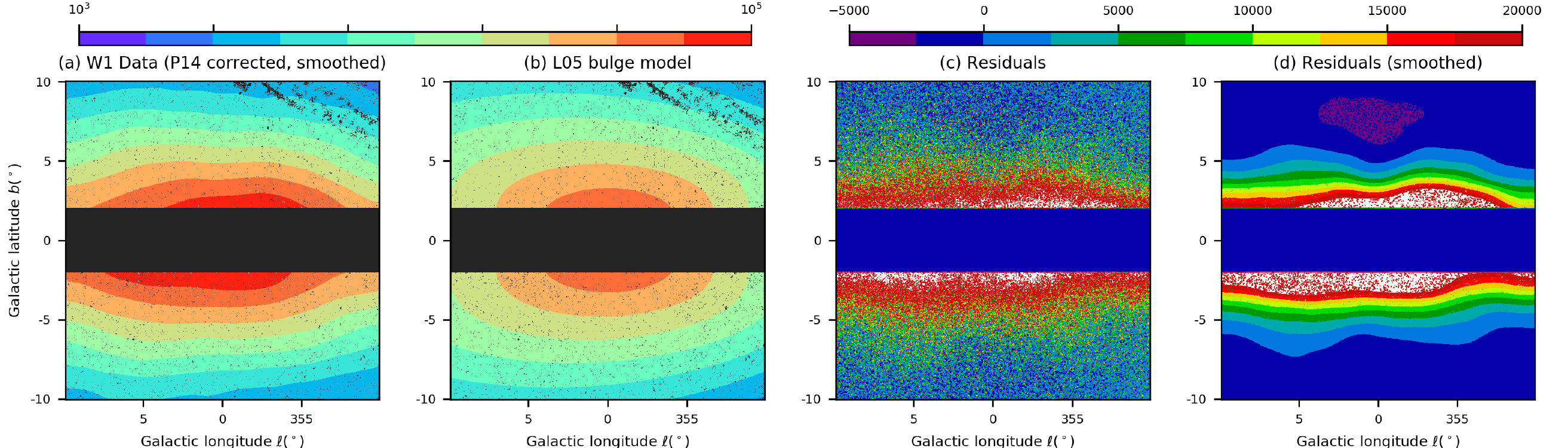}
\caption{Same as Fig. \ref{fig:sfd98-L05}, but with the E(B-V) map of P14. \label{fig:planck-L05}}
\end{figure}

\newpage
Figures \ref{fig:sfd98-L05} and \ref{fig:planck-L05} show the results for the case the boxy bulge model of L05 is subtracted instead of a simple ellipsoidal model from the images corrected with the two different E(B-V) maps, respectively. We can see again from these figures that the residual maps show no obvious X-shape similarly to the previous case. It appears that the difference on the residual map of the Milky Way bulge caused by the two adopted models is not as pronounced as that expected from the toy model in Figure \ref{fig:toymodel}. The residual maps in Figures \ref{fig:sfd98-L05} and \ref{fig:planck-L05} appear more symmetric compared to those in Figures \ref{fig:sfd98-NL16} and \ref{fig:planck-NL16}, which is because the effect of perspective is included in the elongated boxy bulge model substracted.

\section{Discussion} \label{sec:discussion}

\begin{figure}[h!]
\figurenum{8}
\epsscale{0.6}
\plotone{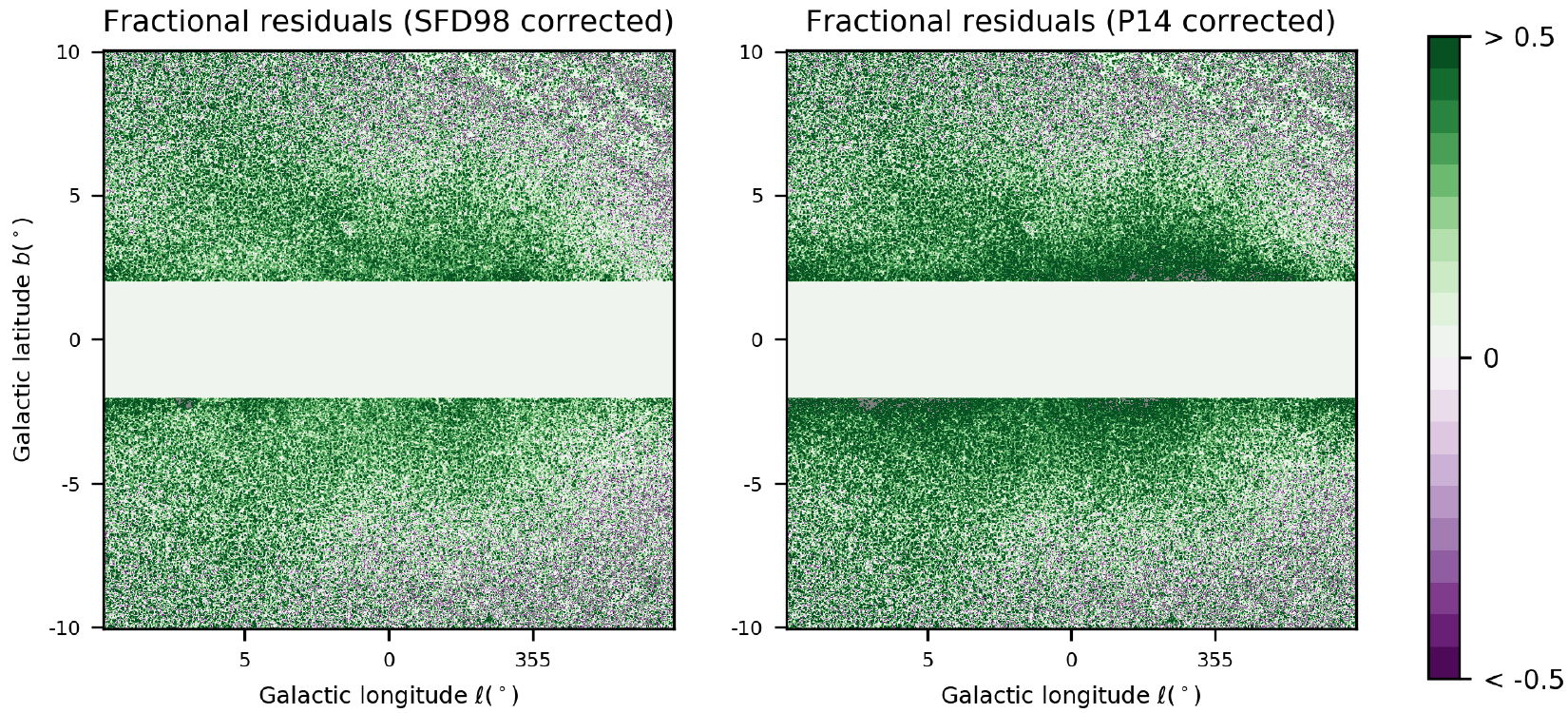}
\caption{Fractional residuals as defined by $(N_{obs}-N_{mod})/N_{obs}$ when the NL16 bulge model is subtracted after the dust extinction correction by the E(B-V) maps of SFD98 (left) and P14 (right), respectively. $N_{obs}$ and $N_{mod}$ are pixel values of the dereddened \textit{WISE W1} data and the bulge model of NL16, respectively. \label{fig:frac-res}}
\end{figure}

We have shown that the analysis in NL16 is affected substantially by the dust extinction correction and partly by a bulge model subtracted from the original image. Therefore, their claim of the direct connection between the X-shaped structure and the double RC phenomenon is seriously questioned as well. However, some faint structures are still remained in our residual maps, and since many of the current studies on the Galactic bulge are based on the two RCs, it is important to check whether they could be related to the double RC phenomenon.

\begin{figure}[h!]
\figurenum{9}
\epsscale{0.5}
\plotone{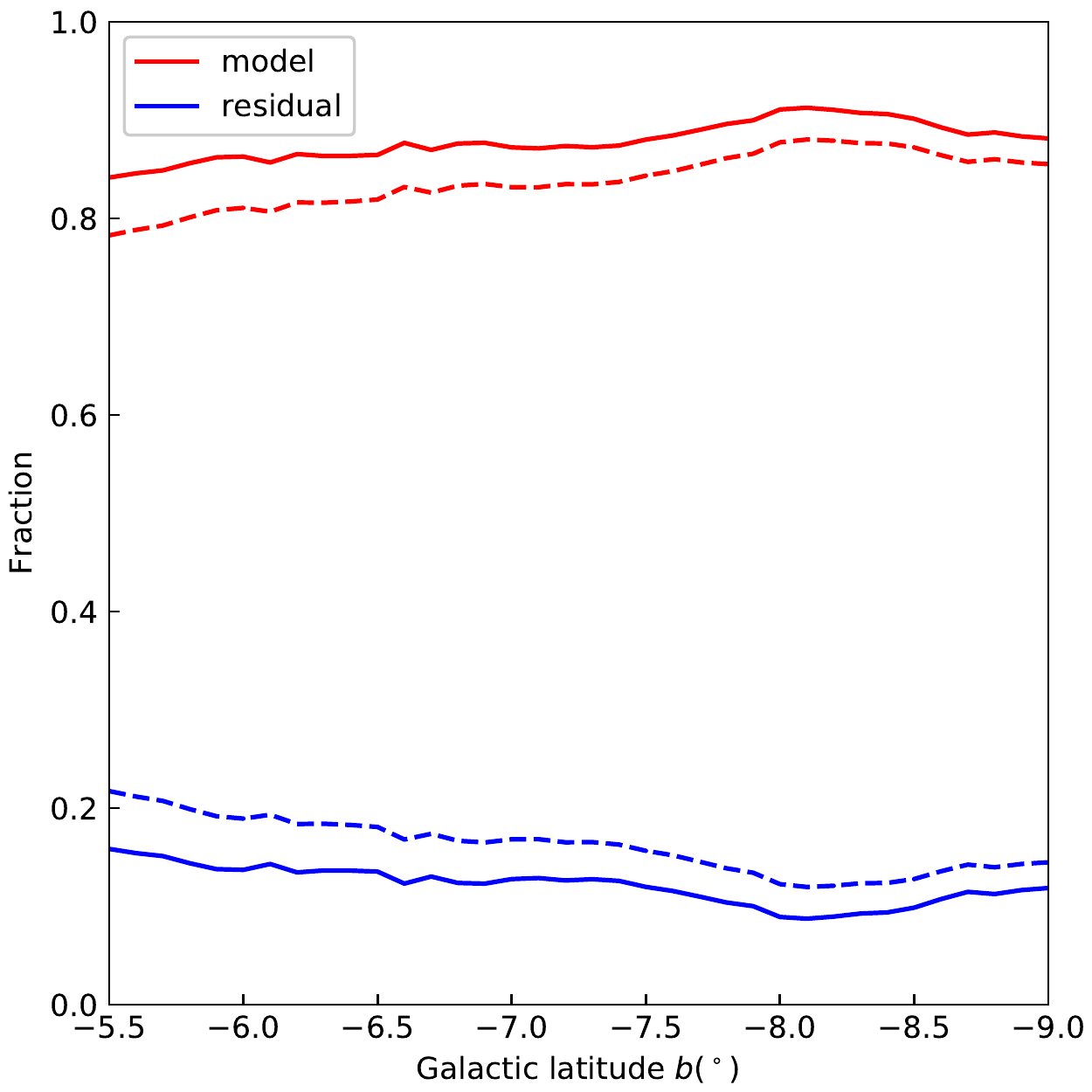}
\caption{The fractions of pixel values of the residual (blue) and NL16 bulge model (red) as functions of Galactic latitude averaged over $-10\degr < l < 10\degr$. Solid and dashed lines are for the E(B-V) maps of SFD98 and P14, respectively. \label{fig:frac-lat}}
\end{figure}

\begin{figure}[h!]
\figurenum{10}
\plotone{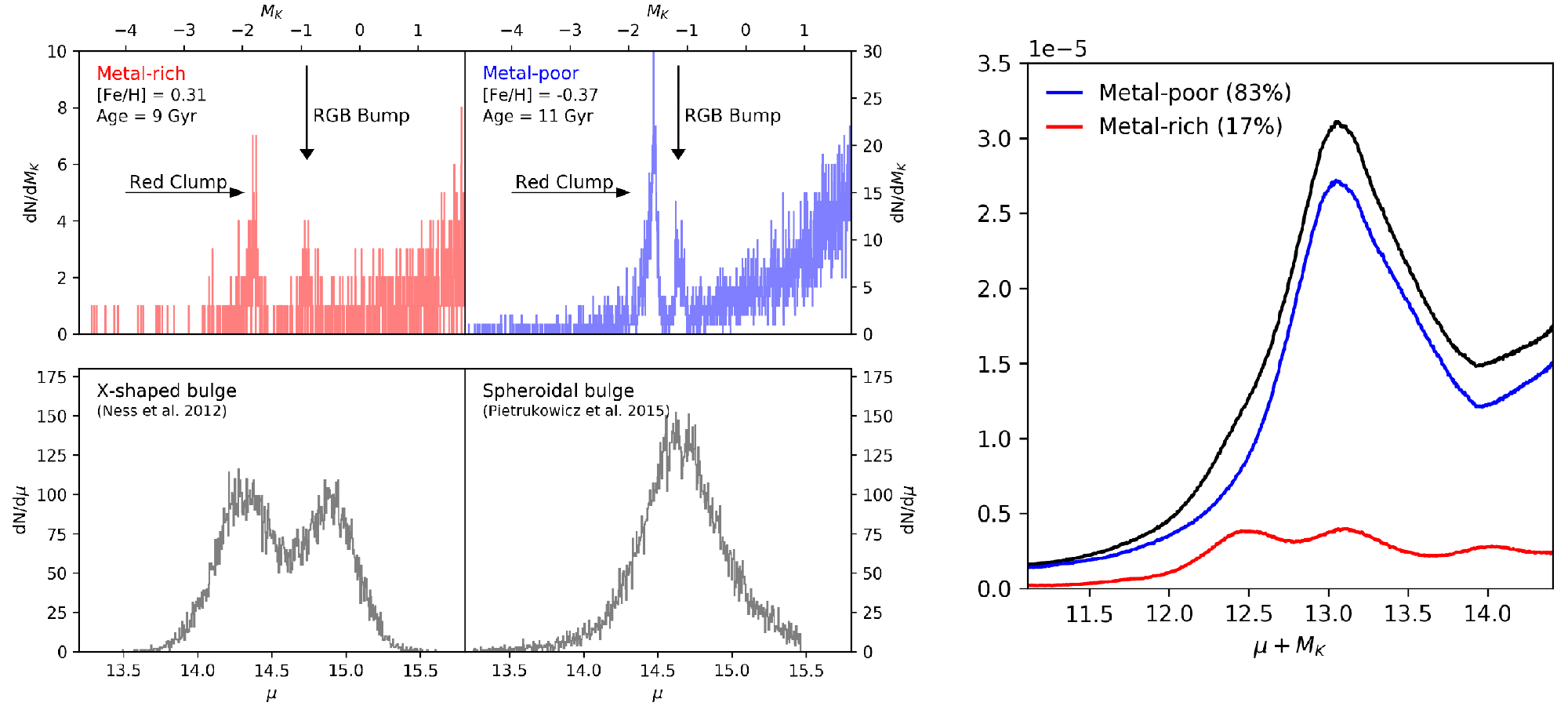}
\caption{Simulations of apparent magnitude distribution for RG + RC stars towards the bulge field at $(l, b) = (0\degr, -8\degr)$. Top: theoretical RG + RC luminosity functions for the metal-rich/poor subpopulations of \citet{2017AA...599...A12} at the same bulge field. Bottom: predicted distance modulus distribution functions for the metal-rich/poor subpopulations under the assumptions that their density maps can be attributed to the X-shaped and spheroidal bulge, respectively. Right: convolved apparent mangitude distribution function. \label{fig:lfunction}}
\vspace{2pt}
\end{figure}

To estimate the level of flux contribution by the structures in our residual maps to the bulge images, we plot in Figure \ref{fig:frac-res} the fractional residuals when the NL16 bulge model is subtracted from the two extinction corrected images considered above. Based on these, we compare in Figure \ref{fig:frac-lat} the fraction of pixel values of the residual maps with that of the NL16 bulge model as a function of Galactic latitude. The level of flux contribution of the residual maps are estimated to be about $15\%$ at $b= -8\degr$, where the double RC is observed to be most prominent. Interestingly, this fraction is very similar to the population ratio ($17\%$) of the metal-rich (MR) stars in the bimodal metallicity distribution function of \citet{2017AA...599...A12} at this bulge field. They have also shown that the density map and kinematics of MR subpopulation are more bar-like, while the metal-poor (MP) subpopulation follows the ellipsoidal density map with a slower rotation. Since the claimed X-shaped structure is supposed to be originated from the Galactic bar, it would be reasonable to assume, arguably, that the stars in this MR subpopulation is connected to the X-shaped structure. Under this assumption, we construct an apparent magnitude distribution for the bulge red giant (RG) + RC populations, in order to investigate whether this minority MR stars could reproduce the observed double RC. The result is shown in Figure \ref{fig:lfunction}, where the theoretical RG + RC luminosity functions are based on the Y2 isochrones \citep{2002APJS...143...499}. For the MR stars in the X-shaped structure, we adopt the distance modulus distribution of \citet{2012APJ...756...22} which is based on the N-body bulge model of \citet{2003MNRAS...341...1179}. For the MP stars, we assume that they follow the triaxial ellipsoidal bulge model of \citet{2015APJ...811...113} obtained from the bulge RR Lyrae survey.

As is clear from Figure \ref{fig:lfunction}, the double peak produced by the X-shaped structure is not revealed in the convolved apparent magnitude distribution function when the majority MP subpopulation with highly concentrated spatial distribution is considered together. As has been suggested by \citet{2015MNRAS...450...L66}, our simulations also suggest that the fraction of at least $45\%$ for the MR subpopulation (stellar component in the X-shaped structure) is required to make a clear double peak in the apparent magnitude distribution function as is observed. Therefore, we conclude that the X-shaped structure claimed by NL16 is not confirmed in this investigation, and even if it is real, the faint structure in the residual map has little to do with the double RC phenomenon.

\acknowledgments
This work was supported by the National Research Foundation of Korea (grants 2017R1A2B3002919 and 2017R1A5A1070354).

\end{document}